\documentclass[preprint,12pt]{elsarticle}

   \usepackage{graphicx}

\usepackage{amssymb,amsmath}

\usepackage{subcaption}

\newcounter{bla}

\journal{Computer Physics Communications}

\begin{document}

\begin{frontmatter}



  \title{A method of calculating bandstructure in real-space with application to all-electron and full potential}


\author[a]{Dongming Li\corref{author}}
\author[b]{James Kestyn}
\author[c]{Eric Polizzi}

\cortext[author] {Corresponding author.\\\textit{E-mail address:} dongmingli@umass.edu}
\address[a]{Department of Electrical and Computer Engineering, University of Massachusetts, Amherst, MA, USA.}
\address[b]{Stellar Science Ltd Co., Albuquerque, NM, USA.}
\address[c]{Department of Electrical and Computer Engineering and the Department of Mathematics and Statistics,
  University of Massachusetts, Amherst, MA, USA.}

\begin{abstract}
  We introduce a practical and efficient approach for calculating the all-electron full potential bandstructure in real space,
  employing a finite element basis. As an alternative to the k-space method, the method involves the self-consistent solution
  of the Kohn-Sham equation within a larger finite system that encloses the unit-cell. It is based on the fact that
  the net potential of the unit-cell converges at a certain radius point. Bandstructure results are then obtained
  by performing non-self-consistent calculations in the Brillouin zone. Numerous numerical experiments demonstrate
  that the obtained valence and conduction bands are in excellent agreement with the pseudopotential k-space method. Moreover,
  we successfully observe the band bending of core electrons.
\end{abstract}


\end{frontmatter}

\section{Introduction}

In solid state physics and electronics, it is crucial to address the issue of obtaining accurate band structure calculations for materials.
One of the earliest band structure calculations was proposed by Cohen in 1960s \cite{Cohen1966} for the silicon crystal.
At the time, the concept of empirical pseudopotential is adopted to simplify the complexity of the electron wavefunction
near the nucleus, and the use of empirical methods further simplifies the process of solving the pseudopotential. In addition, by taking advantage
of the symmetry of the crystal coupled with the introduction of the k-space (Fourier Transformation) method, the domain that needs to be solved
is further reduced to a given unit-cell. A plane-wave expansion scheme is often used to discretize the wavefunctions since it is well-suited to address the periodicity
of the unit-cell.
In the decades following this landmark result and with the advent of high-performance computing,
Kohn-Sham density functional theory (DFT) using ab-initio pseudopotential
has become the cornerstone for the calculation of the band structure of crystals achieving
great success.
However, the k-space pseudopotential method must also relies on other techniques to deal with the complexity of the wavefunctions,
as well as the integration over the energy states.
For example, when the electron wavefunctions
near the nuclei need to be considered, the projector augmented wave technique needs to be used as an alternative \cite{Blochl1994-1,Kresse1999}. In addition,
when calculating the
density of states, it is necessary to use the tetrahedron or the smearing technique to obtain a relatively ideal curve for the density of states \cite{Blochl1994-2}.
The integration over k-space to compute the electron density
needs the so-called special k points in the Brillouin zone to reach numerical convergence \cite{Monkhorst1976,Cances2018}. 

In the context of finite sized atomistic systems, real-space mesh techniques offer a suitable
alternative option to atomic orbitals and plane-wave schemes to discretize the problem (see \cite{Polizzi2020} and references therein).
These techniques allow the quantification of atomic information through the use of universal local mathematical approximations, which
can be systematically refined to achieve convergence at the level of the physical model. Consequently,
they are also well-suited for addressing full-core potential in real-space. As opposed to the use of pseudopotential together with k-space,
full-core potential in real-space permits the computation of
all electron wavefunctions which can be described explicitly over the entire space (including nuclei regions).
Additionally,
real-space mesh techniques like the finite element method, 
can easily handle the treatment of various boundary
conditions including periodic or Bloch-periodic boundary conditions for bandstructure calculations.
In spite of all their benefits, real-space mesh techniques are not frequently employed in k-space bandstructure calculations
due to the challenging handling of long-range Coulomb potentials, which do not converge within a crystal lattice's unit-cell.

This paper proposes a practical and efficient approach
to compute the all-electron bandstructure of periodic solids in
real-space using a finite element method. The issue of long-range divergence is first
resolved by incorporating both the nuclei and electron-electron interaction
potential within a larger finite system that enclosed the unit-cell.
By doing so, we deviate from a fundamental appealing aspect of the k-space method, which depends exclusively
on solving the self-consistent Kohn-Sham system within a relatively compact unit-cell region of the material.
We are able to manage larger system sizes effectively by capitalizing on the recent advancements in solving eigenvalue problems
through high-performance computing solvers like FEAST \cite{Polizzi2009,FEAST}. Additionally, the electron density calculation is easily achieved by integrating
the wavefunctions within finite real-space and finite energy states.
After solving this larger finite system, we can subsequently calculate the periodic wavefunctions
within the unit-cell and determine their respective energy states.
The details of the methodology are provided in Section~\ref{method} while Section~\ref{result} presents various numerical experiments for different
crystal structures to validate the approach.

\section{Methodology}\label{method}


Our starting point is the Density Functional Theory (DFT) and Kohn-Sham one-electron equation described by v-representability
to project in real space the one-electron wavefunction as follows \cite{Kohn1983}:

\begin{equation} \label{eq:dft}
  -\frac{1}{2}\nabla^2\phi(r)+V_{nuclei}(r)\phi(r)+
  V_{KS}(r)\phi(r)=\varepsilon\phi(r),
\end{equation}

where $V_{KS}$ is the Kohn-Sham potential that contains Hartree, Exchange and Correlation terms which are all functionals
of electron density (for clarity, we omit the density functional variable in the Kohn-Sham potential). 
The nuclear potential $V_{nuclei}(r)$ is composed of 
  an infinitely large number of potentials $V_{nuclei}(R_j-r)$ associated with atom $j$ at position $R_j$.

    One possible way of setting up this Hamiltonian consists of filling up the entire space with
    the nucleus potentials first, and then overlays the electron’s interaction potential upon it.
    However, this approach leads to the embarrassment
      of non-convergence of infinite potential energy since the summation of infinite number of long-range Coulomb potentials is divergent.
In the reciprocal k-space space, when using Poisson equation to solve the Hartree and the nucleus potential, the term k=0 must be set to 0 to avoid diverging.
In fact, physically, this is equivalent to treating the unit-cell of a lattice as an electrically neutral one \cite{Komech2015}.
We proceed from this fact to
derive the method presented in this paper.

In the actual process of crystal formation, two atoms are first combined, and then these two atoms combine with another
one to form a three-atom structure, and so on.
As atoms are gradually joined together to form a crystal, the potential energy experienced by an electron within atomic
arrangements remains finite due to its ongoing state of dynamic equilibrium.
Consequently, Electrons are never really experiencing the effect of long-range infinite nuclei potentials, and the 
 potential energy felt by one electron can be better expressed  by
combining $V_{nuclei}$ and $V_{KS}$ in equation~(\ref{eq:dft}) based on each unit-cell, as follows:

\begin{equation} \label{eq:dft_unit}
-\frac{1}{2}\nabla^2\phi(r)+\sum_m^{\infty}\biggl\{\sum_n\biggl[V_{nuclei}^n(R_{m,n}-r)+V_{KS}^n(R_{m,n}-r)\biggr]\biggr\}\phi(r)=\varepsilon\phi(r),
\end{equation}

where $m,n$ correspond to the $m$th unit-cell, and
the $n$th atom in that unit-cell, respectively.
The potentials $V_{nuclei}^n$ and $V_{KS}^n$ are associated with the nuclei and KS potential within an unit-cell (in general, the KS potential
  can be made non-local with the density within the unit-cell).
In equation~(\ref{eq:dft_unit}), we assume that if  $\left|R_{m,n}-r\right|\geq r_0$  where $r_0$ is the average radius
of unit-cell's potential in lattice,
then the second term that represents the net potential will converge to a constant for a given $r$.
Stated otherwise, electrons further from the active area with $r>r_0$,  will not feel the net potential coming from this ``neutral'' unit-cell $m$.
(i.e. the net potential is closed to zero
at a certain distance from the unit-cell).
In the case of polar materials where atoms are positively
or negatively charged, the unit-cell is electrically neutral and the net potential will become conditionally convergent at some point which can be demonstrated
using the Ewald summation \cite{Ewald1921}.
The interaction between unit-cells in crystal lattice is therefore expected to be relatively short-range. 
 We can then assume that the calculation of bandstructure of a lattice only needs to consider the interactions between a certain number of adjacent atoms
 or few unit-cell layers for both non-polar and polar materials.

\begin{figure}[hbtp]
  \begin{subfigure}[b]{0.48\textwidth}\centering
    \includegraphics[width=0.8\linewidth]{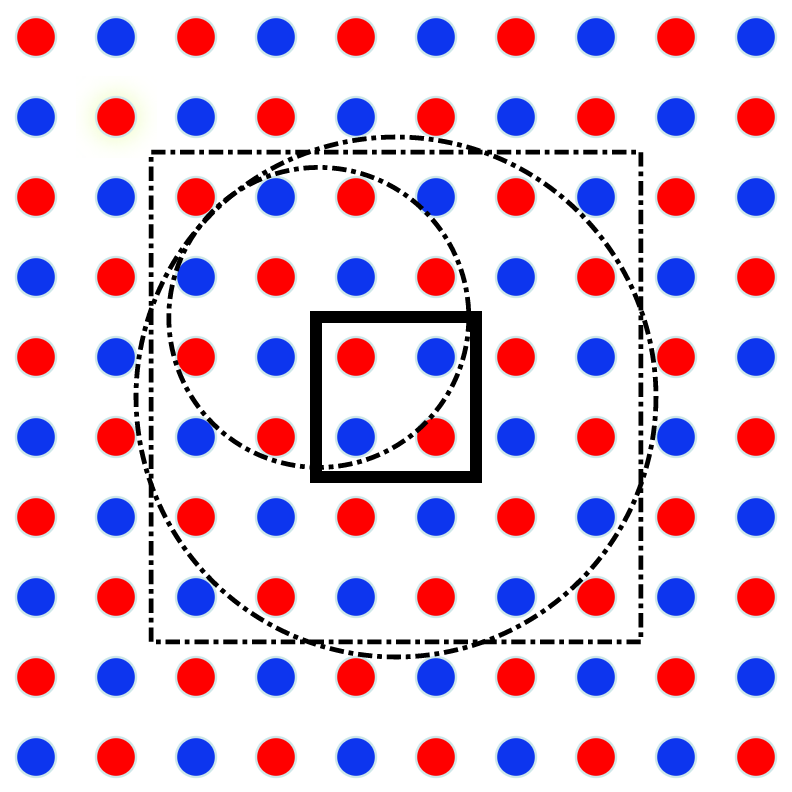}
    \caption{}
    \label{fig12a}
\end{subfigure}
  \begin{subfigure}[b]{0.48\textwidth}\centering
    \includegraphics[width=0.8\linewidth]{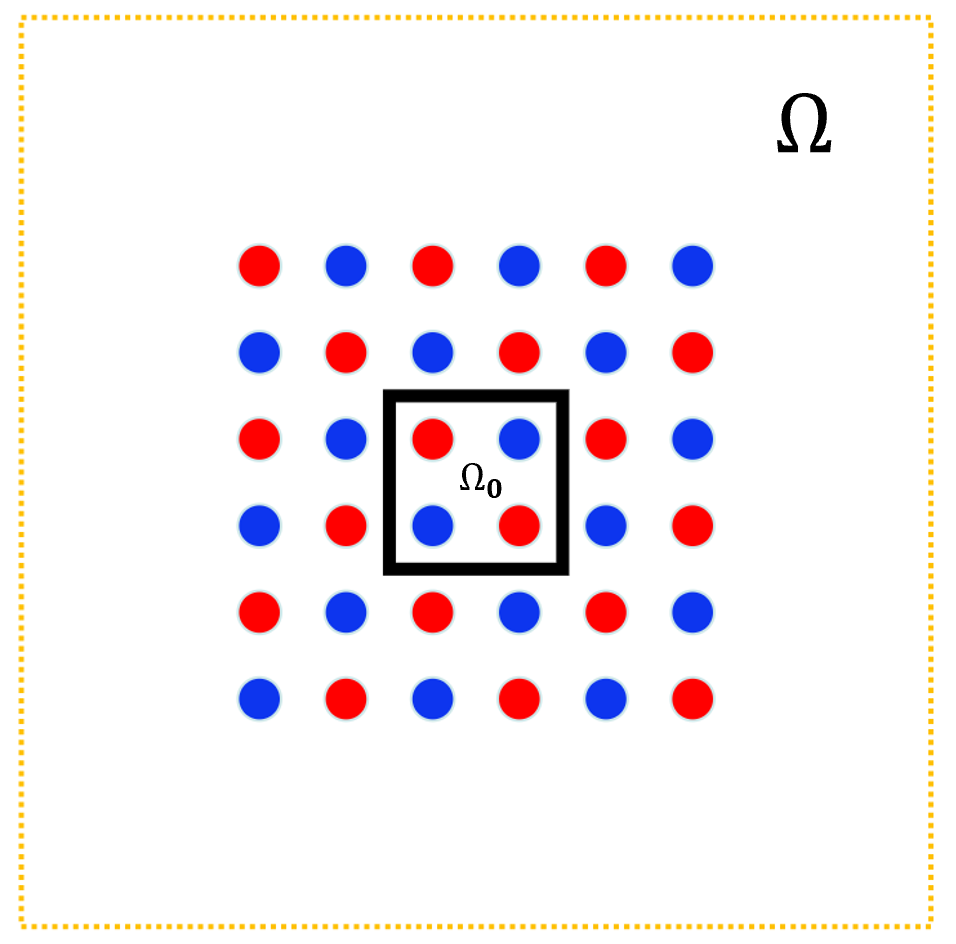}
\caption{}
\label{fig12b}
\end{subfigure}
\caption{(a) Lattice structure with cut-off supercell; (b) A finite isolated system $\Omega$ containing at least one unit-cell $\Omega_0$ and a few adjacent unit-cells.}
\label{fig12}
\end{figure}

   Figure~\ref{fig12a} represents a diagram of how to cut off the supercell based on unit-cell's neutrality. For instance, if the center black box
    is the unit-cell,
    then the cut-off supercell will be the large dashed circle whose size is determined by the small dashed circle with average potential radius  $r_0$ taken at the edge
    of the unit-cell. We note that the value $r_0$ depends on the different types of crystal.
    In practice and for convenience, the outer dashed box can be used as the cut-off supercell instead (even if the atoms
    outside the large dashed circle are included, they will not affect the potential in the inner unit-cell).  Usually,
    for non-polar materials, a supercell containing only 1-2 adjacent layers of atoms is sufficient, while for polar materials,
    a bigger supercell needs to be considered. According to this assumption, the potential term in equation~(\ref{eq:dft_unit}) can be truncated as follows:

\begin{equation} \label{eq:dft_unit_truncated}
  -\frac{1}{2}\nabla^2\phi(r)+\sum_m^{adjacent}\biggl\{\sum_n\biggl[V_{nuclei}^n(R_{m,n}-r) \\+V_{KS}^n(R_{m,n}-r)\biggr]\biggr\}\phi(r)=
  \varepsilon\phi(r).
\end{equation}

While expressing Equation~(\ref{eq:dft_unit_truncated}) in the finite domain $\Omega$ that contains at least one unit-cell ($\Omega_0$ domain) and a few layers of
adjacent atoms (see Figure~\ref{fig12b}),  the problem now consists of solving Equation~(\ref{eq:dft}) self-consistently within $\Omega$:
\begin{equation} \label{eq:dft_omega}
\left[-\frac{1}{2}\nabla^2+V_{nuclei}(r)+V_H(r)+V_X(r)+V_C(r)\right]\phi(r)=\varepsilon\phi(r), \; \; r\in\Omega,
\end{equation}
where the total $V_{KS}$ potential in (\ref{eq:dft}) has been replaced by the Hartree $V_H(r)$, Exchange $V_X(r)$ and Correlation $V_C(r)$ potentials in $\Omega$.

Since $\Omega_0\subset \Omega$, we can extract the converged potentials $V_H(r)$, $V_X(r)$, $V_C(r)$ within the unit-cell $\Omega_0$.
At this point and from the result of the discussion above, the unit-cell potential and electron density should be fully determined and the bandstructure of the lattice
can be retrieved by solving a non-self-consistent Bloch periodic Schr\"odinger equation in $\Omega_0$.
Denoting ${\bf a}$ the 3D primitive vector for the lattice unit-cell, the wavefunction can be written as:
\begin{equation} \label{eq:7}
  \phi(\mathbf{r})=e^{i\mathbf{k}\mathbf{r}}u(\mathbf{r}) \; \; \text{with} \; \;  u({\bf r})=u(\mathbf{r}+m\mathbf{a})
\end{equation}
where $u({\bf r})$ is solution of the following Schr\"odinger equation in $\Omega_0$ using pure periodic boundary conditions:

\begin{equation} \label{eq:Bloch}
\begin{split}
-\frac{1}{2}\nabla^2u(\mathbf{r})-i\mathbf{k}\cdot\nabla u(\mathbf{r})+\frac{1}{2}k^2u(\mathbf{r})+(V_{nuclei}(\mathbf{r})+V_H(\mathbf{r})+V_X(\mathbf{r})+V_C(\mathbf{r}))u(\mathbf{r})\\
=\varepsilon(k) u(\mathbf{r}), \; \mathbf{r}\in\Omega_0.
\end{split}
\end{equation}

Equation~(\ref{eq:Bloch}) is an eigenvalue problem that can be solved for all the relevant {\bf k}-points in the first Brillouin zone
to obtain/plot the bandstructure.

\section{Numerical Experiments}\label{result}

The latest released version of our all-electron finite element NESSIE software \cite{Kestyn2020,NESSIE} was used for performing the Kohn-Sham/DFT/LDA
self-consistent calculations on the finite system~(\ref{eq:dft_omega}). Real space basis sets, like finite element or finite difference,
always involve a large but sparse matrix eigenvalue problem. NESSIE's excellent parallel performance relies on the state-of-the-art
FEAST's eigenvalue algorithm \cite{Polizzi2009,FEAST} for computing all the needed eigenpairs within a given search interval.
It should be noted that the first stage of the proposed approach, which entails solving a finite system in real-space, can be executed using
any discretization schemes (including localized orbitals, plane-wave or any numerical grids).
From the result of the self-consistent procedure, the net potential in the enclosed unit-cell can now be extracted.
We make use of a new finite element mesh to discretize equation (\ref{eq:Bloch}) onto the unit-cell using periodic boundary conditions.
Finite element methods in real space are often used for their reliability, flexibility, and controllable accuracy 
\cite{Lehtovaara2009,Levin2012,Kestyn2020}. They have been introduced by Pask \cite{Pask2005} for solving equation like  (\ref{eq:Bloch})  periodically
using pseudopotential. Finally, without loss of generality, the local density approximation (LDA) was used as the Exchange and Correlation density functional in DFT.

\begin{figure}[htbp]
  \begin{subfigure}[b]{0.48\textwidth}
    \includegraphics[width=\linewidth]{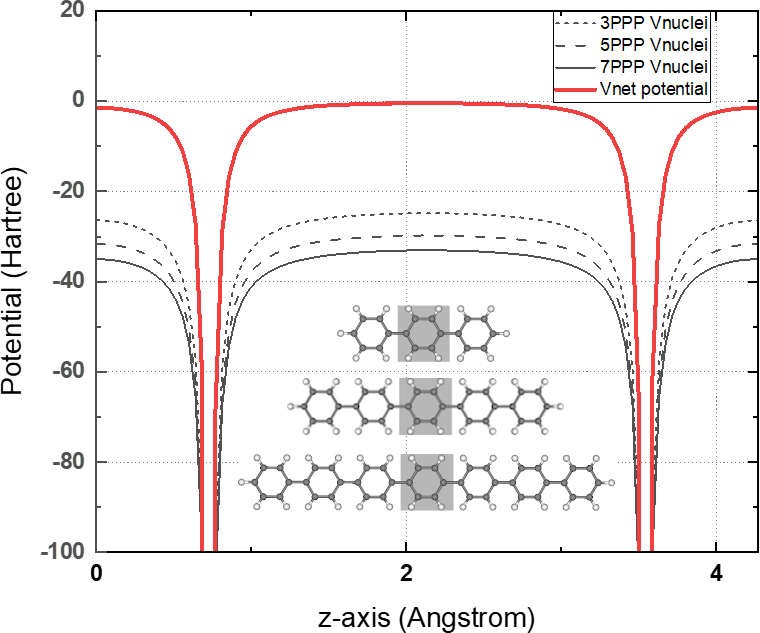}
\caption{}
\label{fig:uca}
\end{subfigure}\hfill
  \begin{subfigure}[b]{0.48\textwidth}
    \includegraphics[width=\linewidth]{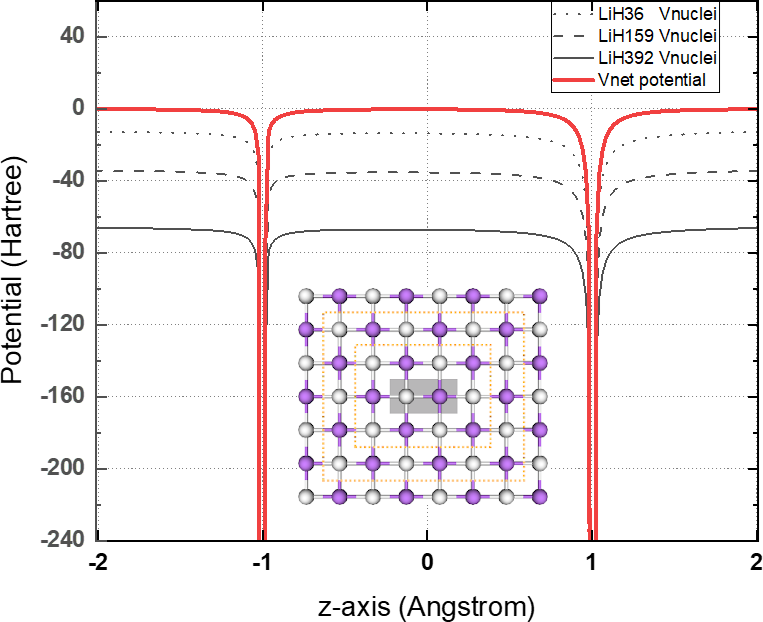}
\caption{}
\label{fig:ucb}
\end{subfigure}
  \caption{Changes in the nuclei potentials and net potentials while increasing the
    numbers of neighboring unit-cells or atoms layers for a given 1D PPP (a) and 3D LiH unit-cell (b).  The potentials are extracted along $z$ at  given $x,y$ points
    which are located at the center of chain for PPP and along the Li-H bond for LiH. All system sizes produce the same net potentials in the plots.
    Only a 2D cross section of the 3D LiH structure is represented for clarity. }
\label{fig:uc}
\end{figure}

Our first set of numerical experiments aim at demonstrating that the net potential in a given unit-cell is expected to only be affected by
few adjacent unit-cells or layers of atoms as it was discussed in the previous Section.
We propose to experiment using both a 1D non-polar poly(p-phenylene) (PPP) polymer, and a 3D polar LiH crystal that uses a cuboid unit-cell.
Fig.~\ref{fig:uca} and Fig.~\ref{fig:ucb} present the variations in the nuclei potentials and
net potentials while increasing the numbers of neighboring unit-cells or atoms layers for a given PPP and LiH unit-cell, respectively.
As previously noted, these results confirm the divergence of the summation of the long-range Coulomb potentials,
while the net potentials (once $V_{KS}$ is included) stay apparently constant for both systems. A more quantitative analysis
shows that one adjacent cell is sufficient for the net potential to converge in the PPP unit-cell to chemical accuracy (close to $10^{-3}$ Hartree),
 while at least two layers of atoms are needed for the net potential to reach convergence in the polar LiH structure. 
Those results help validating our assumption that
a finite system of a certain size is sufficient for solving the energy band of a unit cell in infinite crystal.

\begin{figure}[h]
  \begin{subfigure}[b]{0.32\textwidth}
    \includegraphics[width=0.9\linewidth]{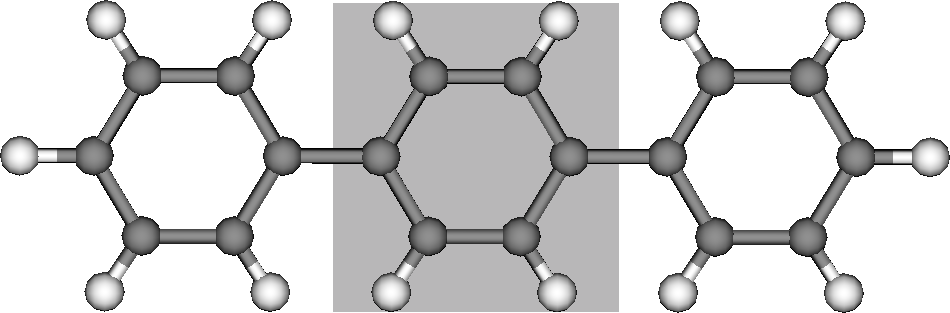}\\[2pt]
     \includegraphics[width=0.885\linewidth]{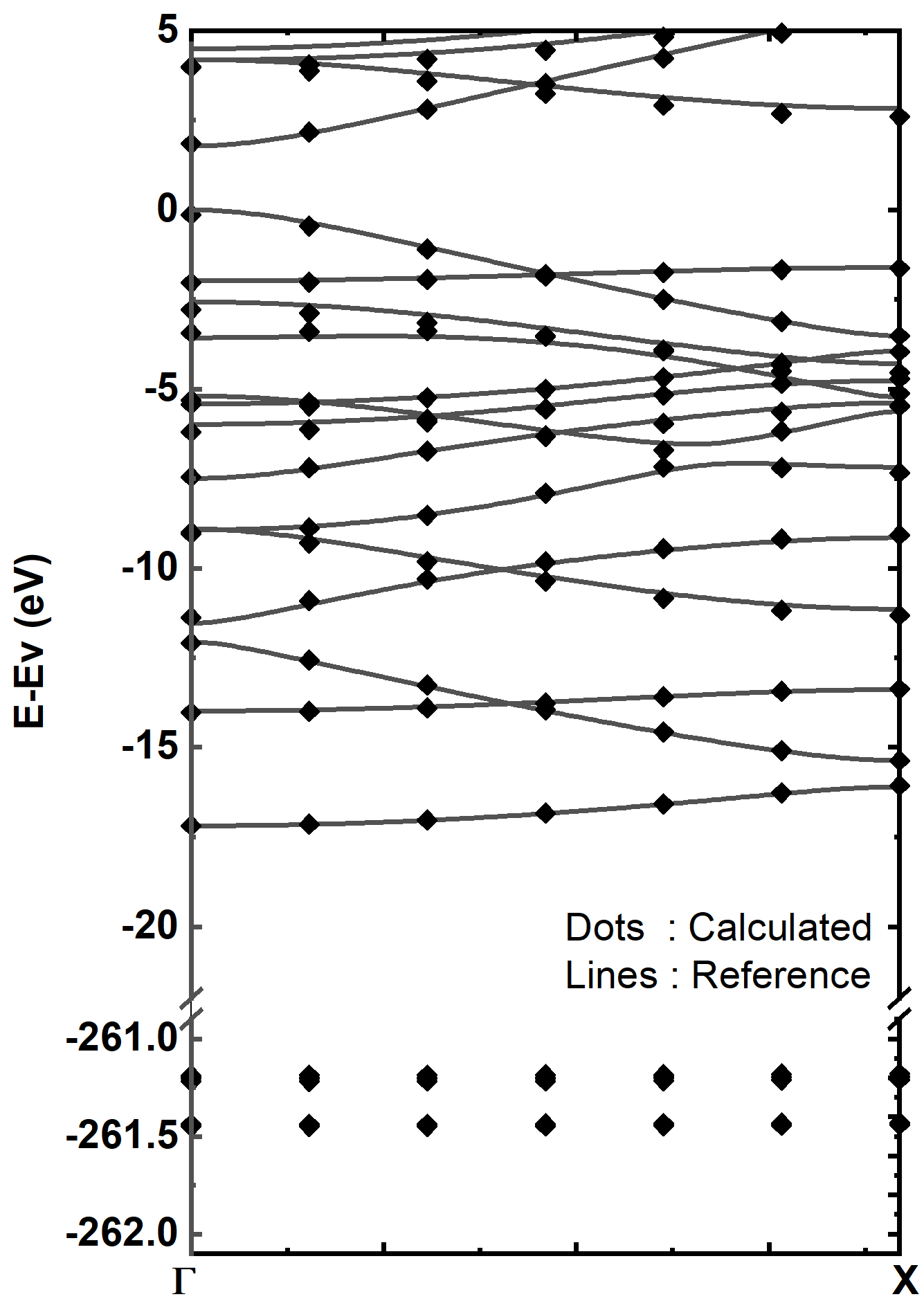}
\caption{}
\label{fig:1dc}
\end{subfigure}
  \begin{subfigure}[b]{0.32\textwidth}
    \includegraphics[width=0.9\linewidth]{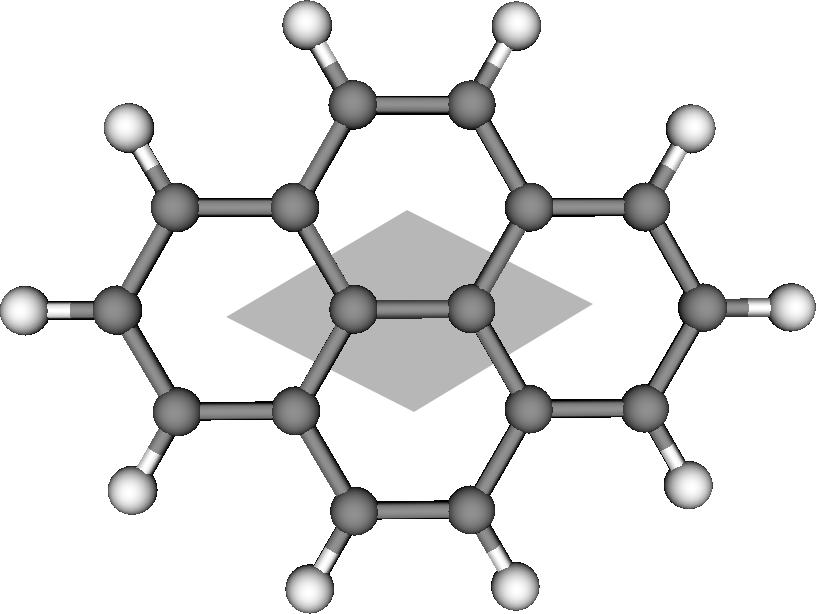}\\[2pt]
     \includegraphics[width=0.9\linewidth]{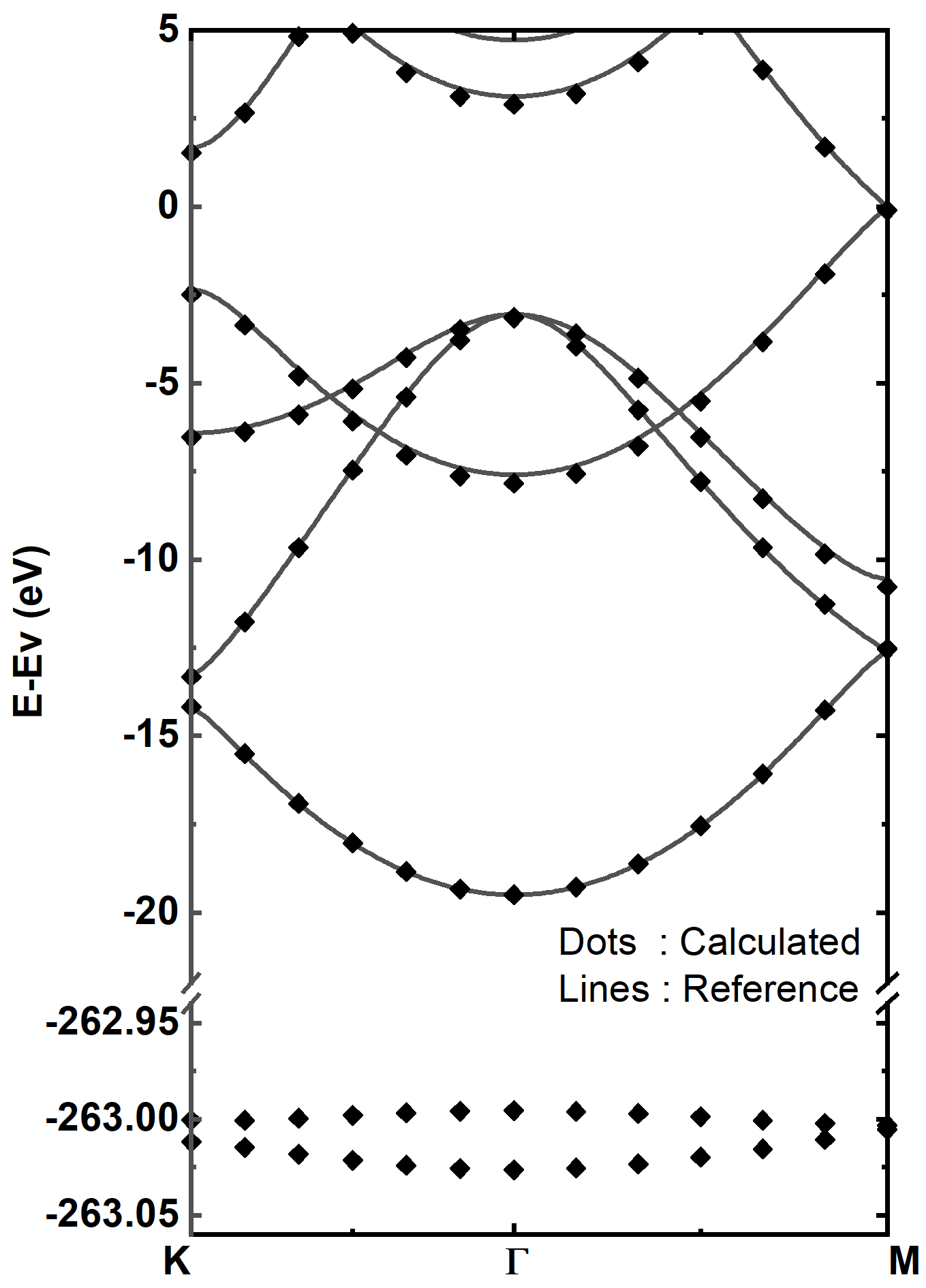}
\caption{}
\label{fig:2dc}
\end{subfigure}
  \begin{subfigure}[b]{0.32\textwidth}
    \includegraphics[width=0.9\linewidth]{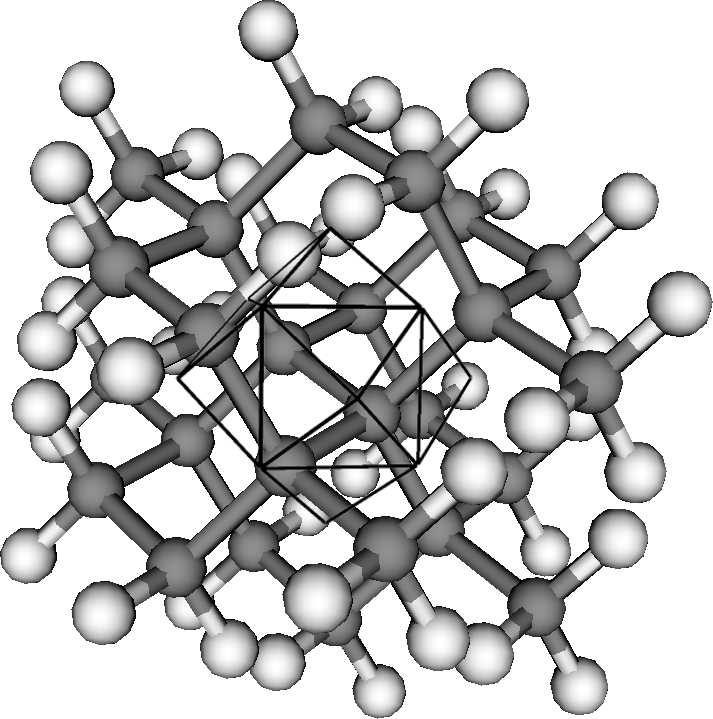}\\[2pt]
     \includegraphics[width=0.895\linewidth]{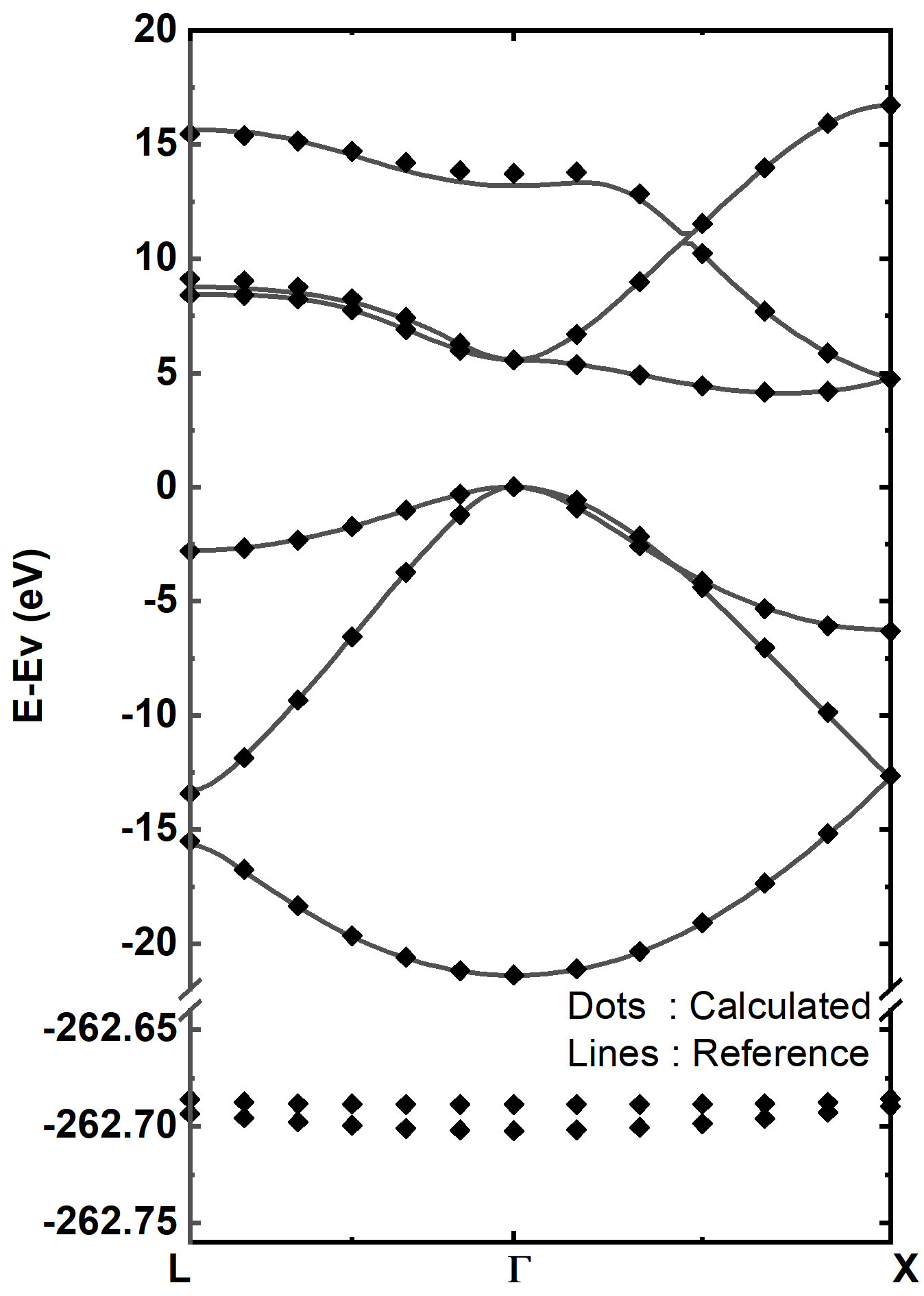}
\caption{}
\label{fig:3dc}
\end{subfigure}
  \caption{Bandstructure calculations for 1D PPP (a), 2D Graphene (b) and 3D carbon Diamond (c).
    Our results for valence and conduction bands are compared with references obtained by k-space pseudopotential calculations.
    Core bands are also presented. The figures on top represent the finite systems that are used to extract the net potentials in
    their corresponding unit-cells.}
\label{fig:c}
\end{figure}

In our second set of numerical experiments presented below, the proposed bandstructure calculation method is applied to
the PPP, graphene, and diamond structures in order to cover the 1D, 2D, and 3D periodic cases, respectively.
\begin{itemize}
\item The PPP unit cell contains 6 carbon and 4 hydrogen atoms,
and it is linked
with other unit-cells at the edges 
to form a 1D periodic chain structure. The bond lengths are: 1.396\AA~  between carbons inside unit cell,
1.1\AA~ between carbon and hydrogen, 1.478\AA~ between carbons connecting unit cells. Angle between carbon-carbon bond and carbon-hydrogen bond is 120$^{\circ}$. 
We use the 3-PPP  unit-cells structure passivated using two hydrogen atoms at the edges, 
as a suitable choice for the finite system to guarantee the convergence of the net potential in the unit-cell (see Fig.~\ref{fig:uca}).
The bandstructure of PPP obtained by solving equation (\ref{eq:Bloch})  within the domain of the unit-cell in the middle, is shown in Fig.~\ref{fig:1dc}.
We note a very good agreement with the result obtained using a more traditional k-space and pseudopotential approach
(computed using the VASP software \cite{Vasp}).
In addition to valence and conduction bands, our all-electron calculation framework makes possible the observation of the core bands. 

\item For 2D periodic structure, we choose graphene as a test material (using 1.42\AA~ for C-C bond length).
  Our finite isolated structure contains 16 carbon atoms and 10 hydrogen atoms used for
  passivation (passivation is necessary to guarantee convergence of the Kohn-Sham self-consistent systems (\ref{eq:dft_omega})).
  A parallelogram including 2 atoms unit-cells (primitive cell) is then used to calculate the bandstructure.  Fig.~\ref{fig:2dc} shows that our
  results for the valence and conduction bands are in very good agreement with the result calculated by the k-space pseudopotential method \cite{Mat}.
  A band bending is also observed from the core electron band. The widths of core and valence bands are about 0.05eV and 20eV, respectively.

\item Diamond is now used for testing 3D periodic features (using 1.542\AA~ for C-C bond length).
  The finite system includes 26 carbon atoms with 42 hydrogen atoms for passivation, while
  2 atoms are used to form a rhombic dodecahedron unit-cell (primitive cell). 
  Fig.~\ref{fig:3dc}  shows the resulting all-electron bandstructure of diamond. Here again, the valence and conduction bands are in good agreement with the result
  calculated by the k-space pseudopotential method \cite{Mat}. The widths of core and valence bands were about 0.017eV and 20eV, respectively. 
\end{itemize}

\begin{figure}[htbp]
    \begin{subfigure}[b]{0.48\textwidth}\centering
    \includegraphics[width=0.6\linewidth]{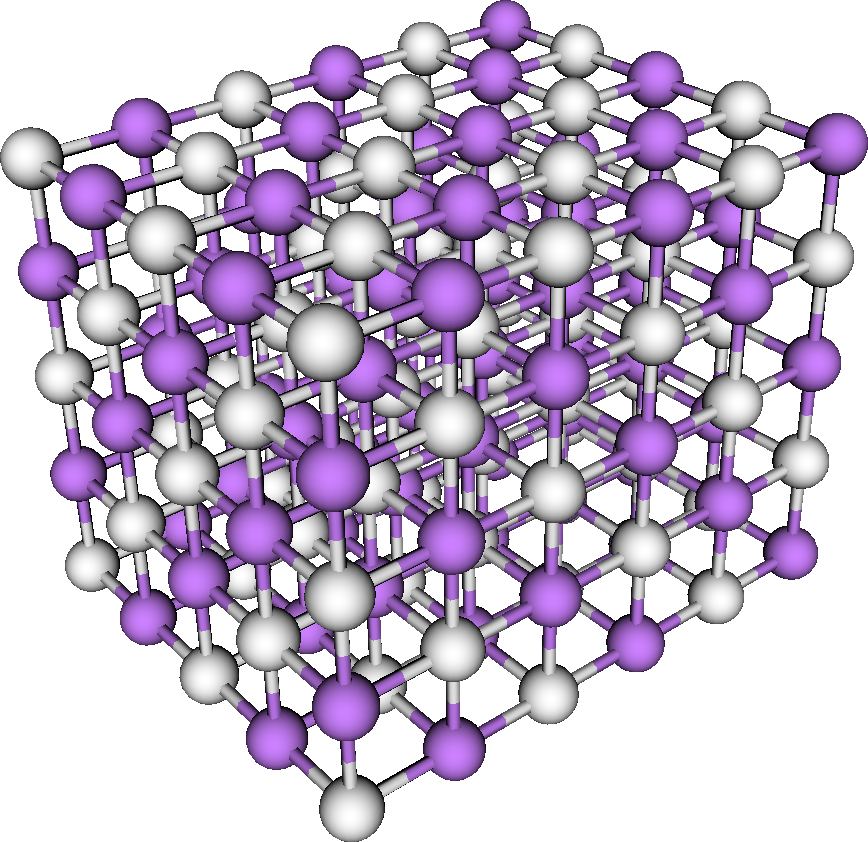}
     \includegraphics[width=0.7\linewidth]{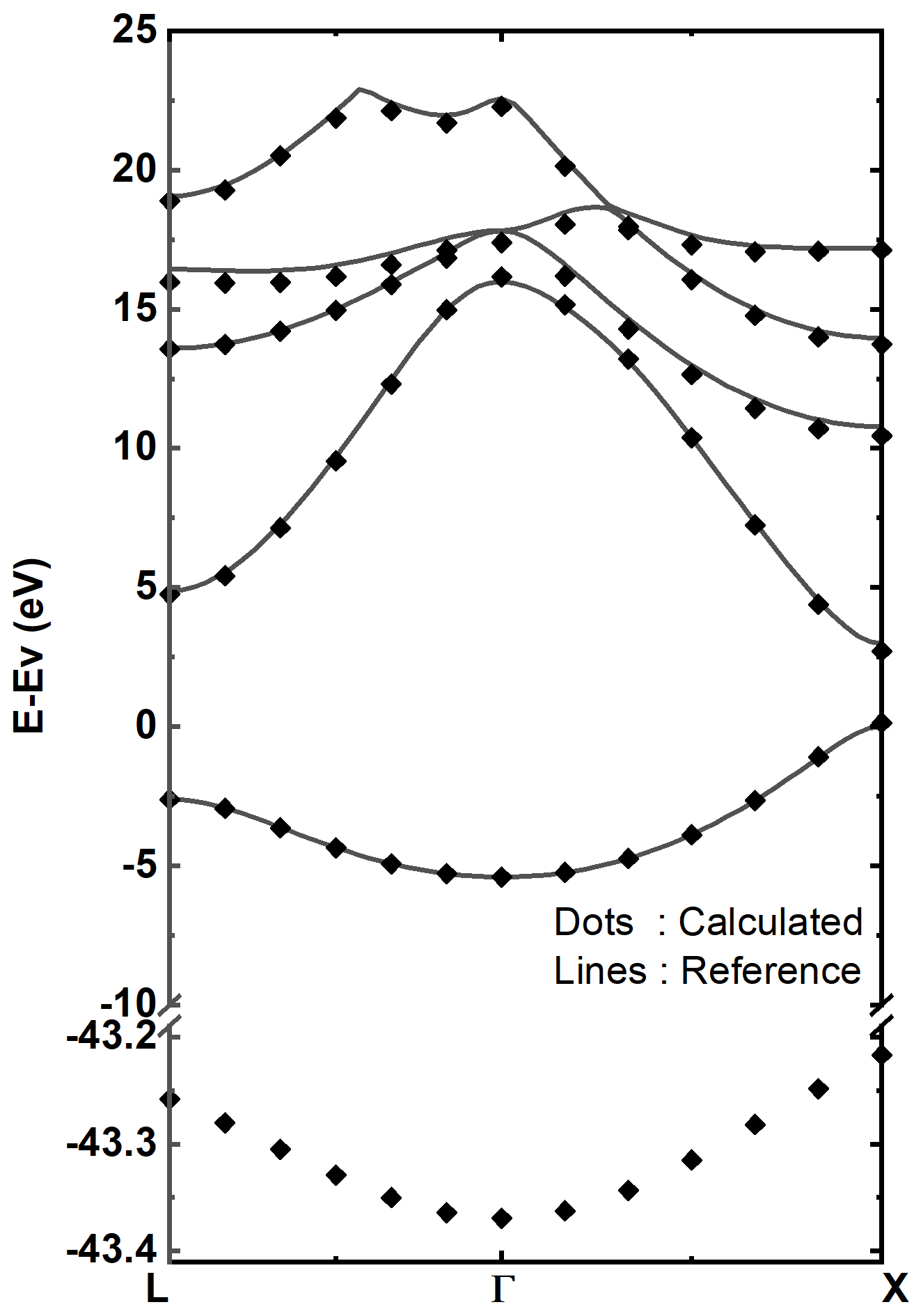}
\caption{}
\label{fig:LiH}
\end{subfigure}
  \begin{subfigure}[b]{0.48\textwidth}\centering
    \includegraphics[width=0.6\linewidth]{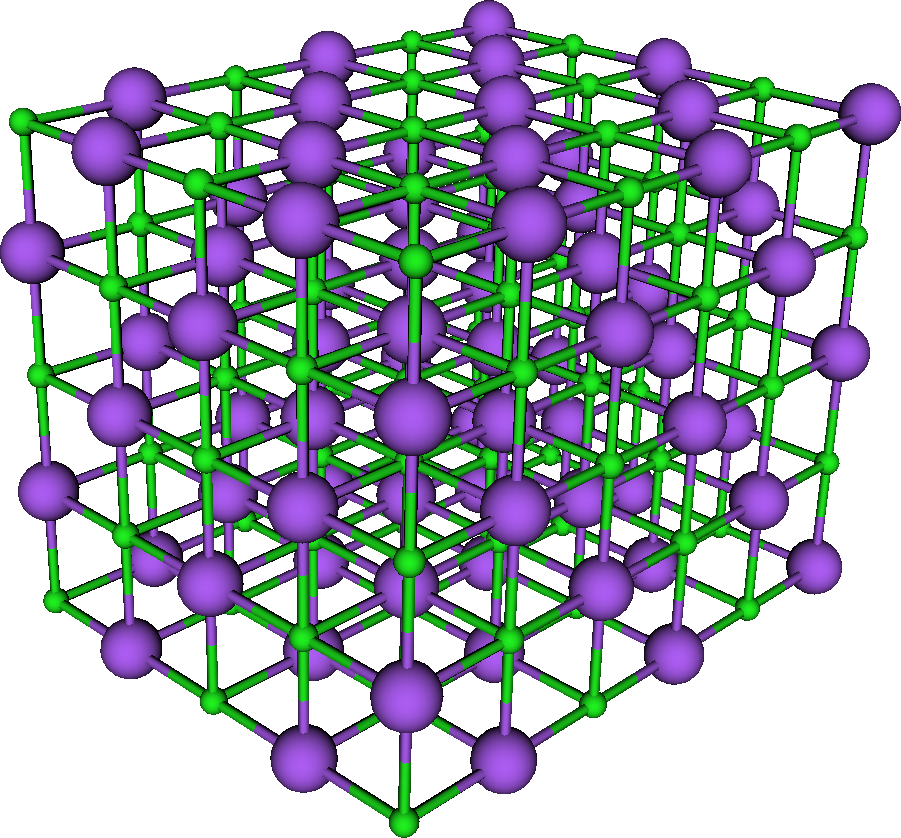}
     \includegraphics[width=0.675\linewidth]{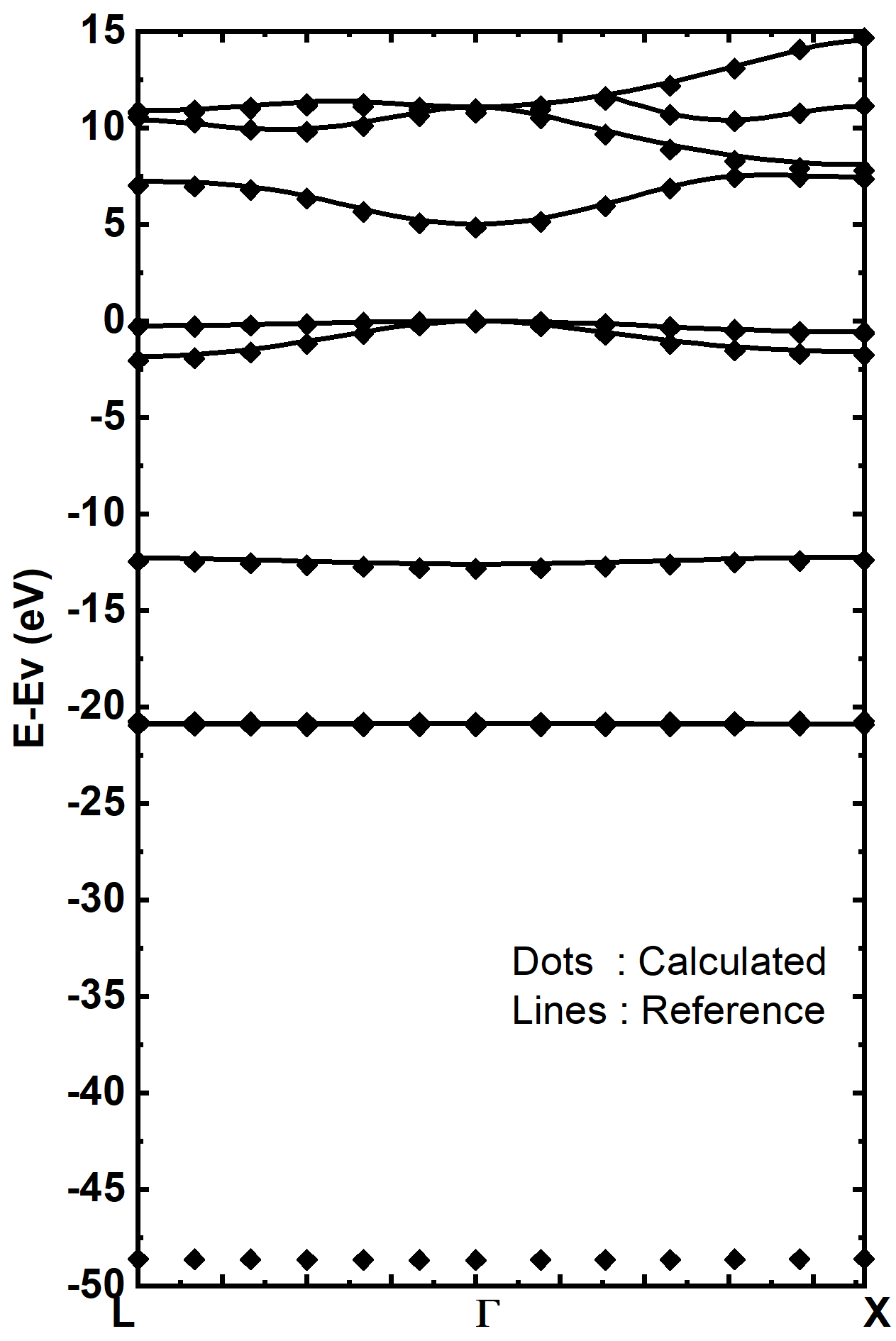}
\caption{}
\label{fig:NaCl}
\end{subfigure}
\caption{Bandstructure calculations for  LiH (a), and NaCl (b).
    Results for valence and conduction bands are compared with references obtained by k-space pseudopotential calculations.
    Lowest core bands for NaCl are not presented. The figures on top represent the finite systems composed of 150 atoms in both cases.}
\label{fig:image2}
\end{figure}

Our third set of numerical experiment is concerned with the 3D polar materials LiH and NaCl. As discussed in Fig.~\ref{fig:ucb} for LiH, two additional layers
of atoms to the central unit-cell, are at least needed to represent the simple cuboid crystal lattice. The finite isolated systems are then composed of 150 atoms for both stuctures (Li-H bond length is  2.01\AA~, and Na-Cl's one is 2.795\AA~).
Fig.~\ref{fig:LiH} and Fig.~\ref{fig:NaCl} show that our results obtained for the valence and conduction bands are in very good agreement with results
from k-space pseudopotential
methods \cite{Mat}. It should be noted that all the results obtained so far for the Carbon-based materials (see Fig~\ref{fig:c}) used P3-FEM (cubic FEM),
but the calculations for LiH and NaCl were performed using P2-FEM (quadratic FEM) to accommodate those larger systems (i.e. improving computing performance).
While the use of P2-FEM is expected to only have a minor influence of the results for the valence and
conduction bands that are relatively high in energy,  P3-FEM would be needed to obtain more accurate core bands (the lowest energy core bands for NaCl are not provided
in Fig~\ref{fig:c}).

\section{Summary}
All electron full potential bandstructure was calculated in real space by taking advantage of the neutrality of atoms in lattice structure.
The action of combining nuclei and electron-electron interaction potential eliminates the divergence of the long-range effect, which enables to extract
the converged net potential in the unit-cell.  Bandstructure calculations can then be performed
non-self-consistently  using Bloch-periodic boundary conditions within the unit-cell to obtain core, valence and conduction bands.
Examples with comparisons were presented for both polar and non-polar materials and multi-dimensional systems. The approach allows
to generalize the domain of applicability of real-space mesh techniques from finite domain to periodic systems, 
and it can be seen as a post-processing procedure for the NESSIE software \cite{NESSIE} which will be soon be integrated as a new feature.
Additionally, and since the Hartree, Exchange, Correlation potential energy of the unit cell are obtained from the isolated finite system,
the vacuum reference is then included. We will discuss this vacuum reference in detail in a subsequent paper.








\bibliographystyle{elsarticle-num}
\bibliography{References}







\end{document}